\documentclass[graybox]{svmult}

\usepackage{mathptmx}       
\usepackage{helvet}         
\usepackage{courier}        
\usepackage{makeidx}         
\usepackage{graphicx}        
\usepackage{multicol}        
\usepackage[bottom]{footmisc} 

\makeindex             

\begin{document}

\title*{Quantum Nanomechanics}
\author{Pritiraj Mohanty}
\institute{Department of Physics, Boston University, 590 Commonwealth Avenue, Boston, MA 02215  \email{mohanty@physics.bu.edu}}
%
%
\maketitle

\abstract{
Quantum Nanomechanics is the emerging field which pertains to the mechanical behavior of nanoscale systems in the quantum domain. Unlike the conventional studies of vibration of molecules and phonons in solids, quantum nanomechanics is defined as the quantum behavior of the entire mechanical structure, including all of its constituents---the atoms, the molecules, the ions, the electrons as well as other excitations. The relevant degrees of freedom of the system are described by macroscopic variables and quantum mechanics in these variables is the essential aspect of quantum nanomechanics. In spite of its obvious importance, however, quantum nanomechanics still awaits proper and complete physical realization. In this article, I provide a conceptual framework for defining quantum nanomechanical systems and their characteristic behaviors, and chart out possible avenues for the experimental realization of {\it bona fide} quantum nanomechanical systems. }

\section{Why Quantum Nanomechanics}
\label{sec:1}

A Quantum Nano-Mechanical (QnM) system is defined as a structure which demonstrates quantum effects in its mechanical motion. This mechanical degrees of freedom involve physical movement of the entire structure. In its current physical realizations, a typical nanomechanical system may consist of 100 million to 100 billion atoms. The mechanical degrees of freedom are therefore described by macroscopic variables. 

Experimental access to the quantum realm is {\it crudely} defined as the regime in which the quantum of energy $hf$ in a resonant mode with frequency $f$ is larger than the thermal energy $k_BT$. The motivation behind this crude definition of the quantum regime is simple. The motion of a QnM system can be described by a harmonic oscillator potential. In the quantum regime, the harmonic oscillator potential energy levels are discrete. In order to observe the effects of discrete energy levels, smearing by thermal energy---due to finite temperature of the QnM system---must be small compared to the energy level spacing, $hf$. However, a formal definition of the quantum regime must involve a proper definition of the QnM system itself, which may include a much more general potential. In any case, the condition $hf \ge k_BT$ gives physically relevant parameters: a nanomechanical structure with a normal mode resonance frequency at 1 GHz will enter the quantum regime below a temperature $T \equiv (h/k_B) f = 48~\mbox{mK}$. Since typical dilution cryostats have a base temperature of 10 mK, nanomechanical structures with frequencies above 1 GHz can enable experimental access to the quantum regime \cite{alexei-PRL,alexei-APL}. The experimental challenge is then to fabricate structures capable of high gigahertz-range resonance frequencies, and to measure their motion at low millikelvin-range temperatures. Because the resonance frequency increases with decreasing system size, one or many of the critical dimensions of the gigahertz-frequency oscillators will be in the sub-micron or nano scale. 

What is the fundamental reason behind a new intiative to physically realize QnM systems in new experiments? Quantum mechanical oscillators have never been realized in engineered structures \cite{bocko}; our physical understanding of quantum harmonic oscillators come from experiments in molecular systems. Furthermore, the obvious extension may also include applications in quantum computing---any quantum system with discrete energy levels and coherence can be construed as quantum bits. Therefore, imagining QnM systems as potential nanomechanical qubits is not farfetched. From a foundational perspective, study of coherence and tunneling effects in any quantum system, somewhat macroscopic in size, lends itself to relevant questions in quantum measurement---usually in a system-environment coupling framework.

Beyond these obvious interests, I argue that a plethora of new and fundamentally important physical problems can be experimentally studied with QnM systems. These problems range from dissipative quantum systems \cite{weiss} and quantum decoherence in the measurement problem \cite{leggett1,leggett2} to phase transition models in condensed matter physics. Furthermore, the structure size of a typical QnM system lies in a regime where the continuum approximation of the elasticity theory is bound to fail \cite{rob-phillips}. The atomistic molecular dynamics approach also becomes severely limited due to the large number of atoms. The size of 100 million to 100 billion atoms requires multi-scale modeling of the elastic properties of QnM systems, which may require novel approaches to computational modeling of large systems. Currently, the state-of-the-art large-scale computing power of a large cluster can handle a size of 100-200 million atoms. Fundamentally, QnM systems may enable a new formalism that marries quantum descriptions of molecules, usually studied in chemistry, with physicist's approach to mechanical systems, quantum or classical.  

This is a list of some of the obvious and not-so-obvious potential applications of QnM systems. Although this list is primarily utilitarian, I argue that uncharted territories bring about unknown concepts. Therefore, it is quite conceivable that---once the experimental activities in QnM systems take off---some yet unknown concept will completely dominate this short list of studies.

\begin{figure}
\centering
\includegraphics[width=0.95\textwidth]{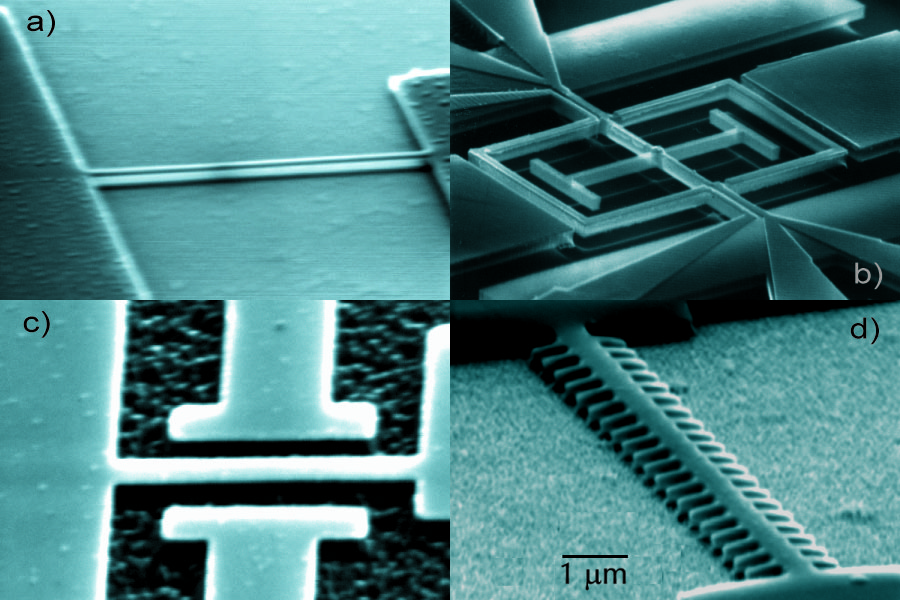}
\caption{Nanomechanical devices important to the foundation of quantum nanomechanics. a) A silicon nanomechanical beam which can work as a nanomechanical memory element by its controlled transition between two nonlinear states \cite{rob-APL,rob-nature}. b) A nanomechanical spin-transport device is used to detect and control spins through a nanowire by the associated spin-transfer torque \cite{spin-PRB}. The device contains a hybrid half-metallic half-ferromagnetic nanowire, which sits on top of a suspended silicon torsion oscillator. c) A nanomechanical beam with electrostatic gate coupling may allow tunable nanomechanical qubit, which will be robust against environmentally induced decoherence due to its macrosopic structure. d) A novel multi-element oscillator structure which allows very high frequency oscillation without compromising detectability of small displacements arising due to high spring constant of a straight-beam oscillator \cite{alexei-PRL}.}
\label{powergraph}
\end{figure}

\section{Quantum Nanomechanical Systems: Definitions and Requirements}

Nanomechanical systems can be defined as mechanical structures free to move in three dimensions with one or many of the critical dimensions under 100 nm. Quantum nanomechanical systems are structures which under certain conditions demonstrate quantum mechanical behavior in their motion. 

\subsection{Dimensionality}

A formal definition of quantum nanomechanics involves quantum mechanics in the acoustic modes of the structure, which include flexural (bending), torsional, and longitudinal modes. These modes represent a geometric change in the shape of the structure \cite{landau,timo}. Therefore an appropriate choice for dimensionality involves how these modes are generated and how they scale as a function of length, width or thickness. In Table 1, we define four distinct dimensions, corresponding to the relative geometric length scales of a rectangular structure. A fundamental distinction among nanomechanical systems with different dimension is the scaling of natural resonant mode frequencies with the natural length scales. For example, in the string limit, resonance frequency varies as $1/L$, whereas in the quasi-1D thin-beam limit, resonance frequency of the natural flexural modes varies as $t/L^2$ according to the elastic theory of continuous media. The relationship between resonance-mode frequencies and geometric parameters also includes a number of relevant material parameters such as material density $\rho$, Young's modulus $Y$, sound velocity $v_s$, and thermal conductivity $\kappa$. 

\begin{table}
\caption{\label{arttype}Dimensionality of nanomechanical systems in terms of geometrical parameters, length $L$, width $w$ and thickness $t$ for rectangular geometry.}
\footnotesize\rm
\begin{tabular*}{\textwidth}{@{}l*{15}{@{\extracolsep{0pt plus12pt}}l}}
\hline\noalign{\smallskip}
Geometrical Parameters&Dimensionality&Description\\
\noalign{\smallskip}\svhline\noalign{\smallskip}
$w,t \ll L$ & 1D & String limit\\
$t \ll w$;\quad $w,t \ll L$ &Quasi-1D&Thin-beam limit\\
$t \ll L,w$&2D&Membrane limit\\
$t\sim w\sim L$&3D&Solid limit\\
\noalign{\smallskip}\svhline\noalign{\smallskip}
\end{tabular*}
\end{table}


\subsection{Classical and Quantum Regimes}

As listed in Table 2, a nanomechanical structure is described by a number of characteristic length scales, important for describing its mechanical motion in either quantum or classical regime. In addition to the scales corresponding to geometry and acoustic phonon wavelength, thermal length defines how far a phonon, the mechanical mode of vibration, extends within the thermal time $\tau_{\beta} = \hbar/k_BT$, where $\tau_{\beta}$ represents the timescale for the system to reach equilibrium with the thermal bath at temperature $T$. The condition for the entire nanomechanical system to be in the quantum regime, the phonon or the mechanical excitation has to extend over the length of the system or $h v_s/k_BT \ge L$. For example, in a silicon nanomechanical beam, the thermal length is $\sim 2~\mbox{micron}$ at a temperature of $\sim$ 100 mK. Therefore, a {\it fully} QnM system of silicon at a temperature above 100 mK should have critical dimensions less than 2 microns in length, irrespective of the quality factor Q.

\begin{table}
\caption{\label{arttype}Length scales of nanomechanical systems in both classical and quantum regime.}
\footnotesize\rm
\begin{tabular*}{\textwidth}{@{}l*{15}{@{\extracolsep{0pt plus12pt}}l}}
\hline\noalign{\smallskip}
Characteristic length scale&Notation&Description\\
\noalign{\smallskip}\svhline\noalign{\smallskip}
geometrical length	     & L,w,t     & Rectangular Structure\\
lattice constant             &   $a$         &  Crystal Structure\\
thermal phonon wavelength    & $\lambda_{th}$         & $\lambda_{th} = h v_s/k_BT$\\
acoustic phonon wavelength   & $\lambda_k$   &    $\lambda_k = 2\pi/k_n$\\
dissipation length           &  $L_d$       &  $L_{d(n)} = v_s \tau_{d(n)} = v_s (2\pi Q_n/\omega_n)$\\
decoherence length           &  $L_\phi$        &  $L_\phi(n) = v_s \tau_\phi(n)$\\
de Broglie wavelength	     &  $\lambda_B$      &          $h/\sqrt{2\pi mk_BT}$ \\
oscillator length            &  $L_O(n)$         & $\sqrt{\hbar/m\omega_n}$ \\
\noalign{\smallskip}\svhline\noalign{\smallskip}
\end{tabular*}
\end{table}

Thermal correlation time is crucial in distinguishing a quantum mechanical system from a classical one, particularly in presence of dissipation (characterized by quality factor Q). Consider, for instance, a temperature at which $\tau_\beta < T_n$, where $T_n = 2\pi /\omega_n$ is the period of oscillation for a given mode of vibration. In this case, correlation between the system and the thermal bath is lost before the end of one cycle of oscillation. Even though, the energy in the classical description is lost in Q cycles, the quantum dynamics is independent from cycle to cycle in this regime. Therefore, the first condition to be in the quantum regime is $\tau_\beta > T_n  = 2\pi /\omega_n$ or $\hbar \omega > k_BT$. For simple harmonic oscillator motion energy level spacing is $\hbar\omega$, so this condition also takes care of the requirement that thermal smearing must be smaller than the energy level spacing.  

Dissipation length is defined through dissipation time $\tau_d = 2\pi Q_n/\omega_n$, where $\omega_n$ and $Q_n$ are the resonance angular frequency and the quality factor for a given mode at a specific temperature. This is the characteristic timescale for loss of energy in the system. In the language of phonons, $1/\tau_d$ is then the inelastic scattering rate of the phonon due to its coupling to the intrinsic or extrinsic environmental degrees of freedom. It is important to compare this to the decoherence of the system at the rate $1/\tau_\phi$. Typically, decoherence of the system can occur much faster than the dissipation of energy, $1/\tau_\phi > 1/\tau_D$ or $\tau_\phi < \tau_D$. 
 
\subsection{Requirement for the Nanomechanical Structures to be Quantum Mechanical}

There are two characterizing aspects of quantum mechanics: coherence and spin. These defining characters manifest in interference effects and statistics, which are not observed in a classical system. Following the spirit of large quantum mechanical systems, one can list two primary requirements for the nanomechanical system to be quantum mechanical. First, quantum coherence of the system must involve proper definition of the relevant physical quantity. Second, macroscopic character of the system must be integrated into the relevant physical quantities to differentiate a quantum nanomechanical system from an idealized point particle in a harmonic oscillator potential.   

In a classical description, the motion of a beam or a cantilever can be  completely described by its transverse displacement at a single point along its length, in particular for flexural or bending motion. Other physical quantities such as velocity and acceleration can be obtained from the transverse displacement $u(x,t)$ for $x \epsilon [0,L]$, where $x$ is the coordinate along the beam axis. Instead of $u(x,t)$, one can define the integrated transverse displacement $\psi$ to describe the beam's motion with a single parameter. $\psi$ can be obtained by integrating the appropriate displacement field $u(x,t)$  along the length with proper boundary conditions. Special cases will involve further constraints on $\psi$ to describe special physical situations (for example, for incompressible beams). $\psi$ can be thought of as an order parameter, representing the motion of a macroscopic structure in both linear and nonlinear regimes. Such a definition can also enable a simple formalism for studying phase transition in the Euler instability region: beyond a critical force a straight beam demonstrates transition to two separate phases of broken symmetry, characterized by mean square displacement \cite{rob-APL}. The second advantage of the definition of an order parameter is the natural connection to the Bose-Einstein Condensation (BEC) description, which contains the essential physics of a large mechanical system, including classical phase transition, macroscopic quantum coherence and multi-stable potential dynamics.
  
\begin{figure}[b]
\sidecaption
\includegraphics[scale=0.5]{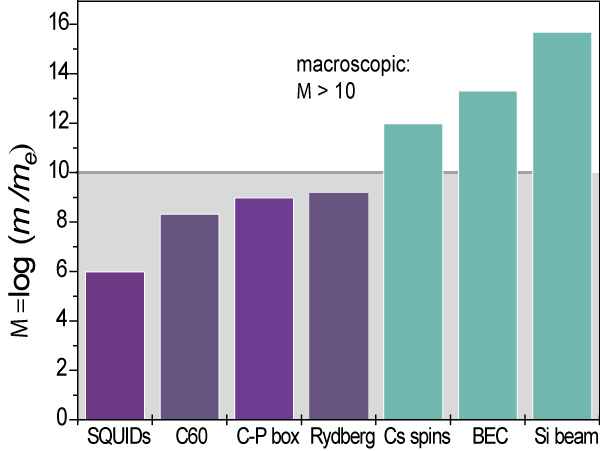}
\caption{Macroscopic nature of the quantum nanomechanical systems shown in comparison with other macroscopic quantum systems by the M-factor, which characterizes the mass of the system relative to the electron mass. With this new definition, structures with an M-factor of 10 or higher will have true macroscopic realism. Quantum nanomechanical systems,  about $\sim$ 10 microns in size, can in fact be seen by naked eye.
}
\end{figure}

The quantum mechanics of a nanomechanical system can be described by the order parameter with an amplitude and a phase: $\psi = |\psi|e^{i\phi}$. In the quantum regime, the nanomechanical system becomes a phase coherent system with a ``macroscopic" quantum wave function. Since, the quantum motion of nanomechanical systems involve matter waves, the macroscopic nature of the structure can simply be  defined in terms of its mass $m$ through a new quantity, M-factor, which is defined by $M = log (m/m_e)$, where $m_e$ represents the mass of an electron. The concept behind this simple definition is two-fold. First, if the M-factor is larger than 10, then it can be considered truly macroscopic from the perspective of our experience in the ``everyday world." Figure 2 displays the M-factor for a number of macroscopic quantum systems. Nanomechanical structures in the quantum regime can contain about a billion atoms, and their size in the range of microns can in fact allow them to be observed by naked eye. The second reason is to emphasize the mass of the system in determining its quantum mechanical behavior in terms of coherent matter waves.

Coherence of the matter wave, representing quantum nanomechanical motion, can be characterized by a decoherence time $\tau_\phi$. In simple cases, the decoherence time or the associated decoherence length may be dominated by the de Broglie wavelength, which describes the spread of a gaussian wave packet. However, a proper analysis of intrinsic decoherence mechanisms must be made for the correct estimate. The simplest approach is to follow the convention in defining decoherence of Schrodinger cat states in BEC \cite{dalvit,moore}. 
       
\section{Potential Quantum Nanomechanical Systems}

It is necessary for the physical realization of quantum nanomechanics that the appropriate conditions of quantum mechanics are satisfied. As mentioned earlier, the requirement of high resonance-mode frequency at low temperature, $\hbar\omega > k_BT$, may not be sufficient for the system to be fully quantum mechanical. The fundamental difficulty is to legislate what time scale or corresponding length scale, among those listed in Table 2, is the single characteristic length scale that determines if the macroscopic nanomechanical system is in the true quantum regime. Although there are relevant conventions in both atomic bose condensation and electronic mesoscopic physics, it is important to obtain experimental data to be able to fully identify the appropriate length and time scales. In this section, I list four different classes of experimental nanomechanical systems in which efforts are currently being made towards the observation of quantum effects. 


Linear displacement and velocity of the nanomechanical systems can be detected by a number of transduction mechanisms, which allow conversion of a mechanical signal to an electrical signal. These include electrostatic detection technique in which the beam's motion is detected by measuring the change in the capacitance between an electrode on the beam and a nearby control electrode. As the distance between the two plates changes, the capacitance changes. In order to induce motion in the beam, an electric field can be applied between the two plates at or near the resonance frequency of the beam. In the optical technique, beam's displacement can be measured either directly or through an interferometric method. Because of the millikelvin temperature requirement it is difficult to employ optical techniques, as the minimum incident power from the laser will tend to increase the temperature substantially. The electrostatic technique is unsuitable because of the large parasitic capacitance between the different parts of the device and the surroundings. A variation of the electrostatic method is the coupled-SET (Single-Electron Transistor) technique in which the change in the capacitance between the two electrodes due to the motion of the beam is detected by a single-electron transistor. In this configuration, one of the electrode plates is used as a gate of the SET transistor. The change in the gate voltage is measured by detecting the change in the source-drain current of the SET. In spite of its sophistication, it is difficult with this technique to detect gigahertz-range motion in a straight beam, as the change in capacitance generated by the motion at these frequencies is very small.

\subsection{Straight-Beam Oscillators}

A straightforward approach to the quantum regime involves measurement of displacement or energy of a straight-beam nanomechanical structure in the gigahertz range at a temperature $k_BT < hf$. Although such submicron structures with expected gigahertz-range frequencies are now routinely fabricated in laboratories, motion at frequencies in the gigahertz range has not been detected with equal ease.   

The fundamental problem in straightforward miniaturization of beam or cantilever oscillators is the increase in the stiffness constant along with the increasing frequency, which is required for getting into the quantum regime. For a straight beam in the thin-beam approximation, stiffness constant increases as $w (t/L)^3$, or $1/L^3$ if the cross-sectional dimensions $w$ and $t$ are kept constant. A high spring constant, typically in the range of 1000--10000 N/m, results in undetectably small displacements, typically in the range of 1--10 fm (femtometer), corresponding to a force of 1 pN.  However, experimental considerations such as nonlinearity and heating require the range of force to be even smaller than that. Therefore, the problem of detecting motion in the quantum regime translates to the problem of detecting femtometer-level displacements at gigahertz frequencies, assuming that the structure cools to millikelvin-range temperature. In straight-beam oscillators, thermal phonon wavelength $\lambda_{th}$ becomes orders of magnitude larger than the cross-sectional dimensions, which prevents the central part of the beam from cooling to the required millikelvin temperature.


\subsection{Multi-Element Oscillators}

Design of structures for the detection of quantum motion at gigahertz frequencies therefore is a two-fold problem. First, the normal-mode frequencies have to be in the gigahertz range. Second, the structure in these gigahertz modes must have a much lower spring constant ``$k_{eff}$" to generate a detectable displacement or velocity. This cannot be achieved with simple beams as ``$k_{eff}$" and ``$\omega $" are coupled by trivial dispersion relations. The problem is to find a structure with certain modes in which ``$k_{eff}$" and ``$\omega $" can be decoupled. However, decoupling of ``$k_{eff}$" and ``$\omega $" cannot be achieved in single-element structures.  


One type of multi-element structure, comprising of two coupled but distinct components \cite{alexei-PRL} has been experimentally studied. Small identical paddles serve as the frequency-determining elements, which generate gigahertz-range natural frequencies because of their sub-micron dimensions. The paddles are arranged in two symmetric arrays on both sides of a central beam, which acts as the displacement-determining element.  Because of its multipart design, the structure displays many normal modes of vibration, including the fundamental mode and numerous complex modes. By design, there exists a class of collective modes at high frequencies, apart from all other normal modes. In the collective modes, the sub-micron paddles move in phase to induce relatively large amplitude of motion along the central beam at the same frequency.

In recent experiments, the antenna structure has been studied in detail at low temperatures by the magnetomotive technique \cite{alexei-APL,alexei-PRL}. It exhibits the expected classical behavior at the low frequency modes. A class of high-frequency collective modes are observed in the range of 480 MHz---3 GHz. At temperatures corresponding to high thermal occupation number $k_BT/hf$, the high-frequency gigahertz modes show the expected classical behavior, equivalent to the linear Hooke's law. 

In the quantum regime, $N_{th} \equiv k_BT/hf \le 1$, the gigahertz modes show discrete transitions in contrast to the classical behavior of the same modes at higher temperatures, $N_{th} \gg 1$. The 1.5-GHz mode at 110 mK ($N_{th}\rightarrow 1$) displays discrete transitions as a function of driving force or magnetic field \cite{alexei-PRL}.  While the transitions do not always occur at exactly the same field values from sweep to sweep, the jump size remains unchanged, suggesting that the oscillator switches between two well-defined states. Although these reproducible discrete jumps could indicate transition to quantum behavior, it is difficult to gain more insight into the nature of the two states from the data.

A higher frequency resonance mode at 1.88 GHz was studied down to a cryostat temperature of 60 mK \cite{alexei-new}, deeper in the quantum regime, corresponding to $N_{th} \sim 0.66$.  Figure 3 shows a four-state discrete velocity response in the form of a staircase as a function of continuous driving force.  In frequency domain, the response displays clear gaps in the growth of the resonance peak as the magnetic field is increased in equal increments of 0.2 tesla. The lineshape remains almost Lorentzian, which excludes the possibility of a standard classical nonlinear effect. 

The response is highly reproducible with a strong temperature dependence.  Above the mixing-chamber temperature of 1.2 K, corresponding to $N_{th} \sim 14 $, the discrete transitions abruptly disappear, and the mode response varies smoothly with applied drive, consistent with a classical behavior. In addition, low megahertz frequency modes of the same oscillator at the lowest measured refrigerator temperature of 60 millikelvin demonstrate expected classical dependence. Further investigations are currently underway to fully characterize this remarkable effect.

\begin{figure}
\centering
\includegraphics[width=0.95\textwidth]{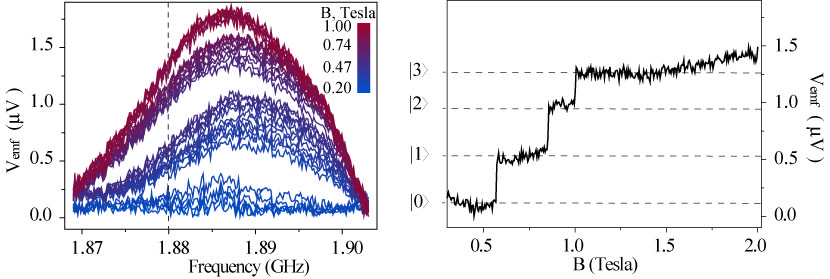}
\caption{Mechanical response of the antenna structure in the quantum regime. a) Amplitude response of the 1.88-GHz mode at a (cryostat) temperature of 60 mK, corresponding to a thermal occupation number $N_{th} \sim 0.66$, demonstrates gaps as a function of increasing driving energy. b) A continuous sweep of the driving force (provided by magnetic field) at a single frequency 1.88 GHz shows discrete jumps.  }
\end{figure}

\subsection{Tunneling Two-State Oscillator in a Double-Well Potential}

A different approach to the experimental realization of quantum effects in the motion of a nanomechanical system involves quantum tunneling of the entire structure between two physically distinct states. A nanomechanical beam oscillator can be driven in the nonlinear regime where the equation of motion is given by the standard Duffing oscillator expression. The amplitude response of the oscillator changes from the standard Lorentzian form in the linear regime to an asymmetric hysteretic form in the Duffing regime where the oscillation amplitude is multi-valued. Hysteresis is demonstrated in frequency sweep, as the oscillator follows two different states depending on whether frequency is swept forward or backward through the bistable regime.    

Recent experiments have demonstrated controlled switching between the two nonlinear states of the oscillator in the classical regime \cite{rob-APL}. The addition of a slowly-varying driving force modulates the double-well potential, associated with nonlinearity, and enables the system to go over the potential barrier following the sub-threshold modulation signal. This classically coherent transition between the two states can also be enhanced by the application of white noise to the system, which results in stochastic resonance for a given range of noise power \cite{rob-nature}. As temperature is decreased, another pathway for transition between the two states opens up as the probability for the system to directly tunnel through the barrier increases. Quantum mechanical tunneling of the macroscopic nanoscale oscillator is a fundamentally different realization of quantum nanomechanics. Towards this end, there have been a few theoretical analyses \cite{lawrence,peano,zwerger}. However, new experiments are clearly needed for a better definition of this problem.

\subsection{Coupled Nano-Electro-Mechanical Systems}

Currently, most of the theoretical activities have been focused on coupled-NEMS (nano-electro-mechanical systems). In this setup, motion of a nanomechanical resonator is analyzed through its coupling to a measurement appartus provided by either a single electron transistor (SET) or a cooper-pair box. Recently, two experimental groups have demonstrated that it is possible to detect mechanical motion of oscillators using the SET-detection technique \cite{cleland,schwab}. Although the initial experiments (performed on oscillators with relatively low frequencies in the megahertz range, and hence high thermal occupation number $N_{th}$) have demonstrated relatively high displacement detection sensitivity, it is not clear if at higher gigahertz-range frequencies the detection sensitivity will be as high. Nevertheless, this technique, motivated essentially by qubit experiments in mesoscopic physics, offers the possibility of detecting quantum motion in straight-beam oscillators.      

\section{Endnote} 

The field of quantum nanomechanics is off to a good start. However, new experiments are needed to build a solid phenomenology since this is a new territory in respect to our intuition of large quantum systems. It is still not known if  quantum mechanics in its current form remains a valid description of systems as large as the nanomechanical structures with billions of atoms \cite{leggett2}. There are further complications due to finite dissipation and finite decoherence. Therefore, it is important to manage our expectation of what we ought to observe in experiments. Perhaps it is prudent to be guided by phenomenology as we build up the conceptual and theoretical framework. 

Experiments with fundamentally different measurements of displacement, velocity, acceleration and other mechanical properties are needed to be performed on nanomechanical systems with high frequencies at low temperatures. It is important to include multiple actuation and detection approaches to address the issues of eigen selection, quantum non-demolition and back action.  Materials choice also becomes an important concern, hence repetition of the same measurements on structures of different materials can also elucidate fundamental quantum effects. Beyond the obvious materials such as silicon, silicon carbide, gallium arsenide, carbon nanotube and carbon 60, it is important to explore doped and undoped diamond \cite{imboden}, aluminum nitride, graphene \cite{cornell} and other new materials for nanomechanics.

On the theoretical side, a bootstrap approach to developing the appropriate framework will be needed. Proper definitions of relevant length scales and time scales must be done along new experimental data. Furthermore, basic theoretical analyses must be done to understand the classical dynamics and the energy spectrum of extended mechanical objects of relevant geometry.  A lot of progress is being currently made in multi-element oscillators, where exact calculations of the energy spectrum for driven and undriven periodic structures have been done \cite{jerome}.  Similar calculations are needed to define and understand measurable quantum properties of realistic systems studied in experiments. Lastly, it is important to continue the debates and discussions \cite{comments} which are fundamentally important to developing a new field. 

\begin{acknowledgement}
The experimental part of the work has been done in collaboration with numerous members of my group, including Alexei Gaidarzhy, Robert Badzey, Guiti Zolfagharkhani, Yu Chen, Matthias Imboden, Josef-Stefan Wenzler, Tyler Dunn and Diego Guerra. This work is supported by National Science Foundation (grant no. DMR-0449670).

\end{acknowledgement}
%

%
%

\end{document}